\documentclass{aa}  

\usepackage{color}
\bibpunct{(}{)}{;}{a}{}{,}
\usepackage{graphicx}
\usepackage{txfonts}
\usepackage{lscape}             
\usepackage{placeins}     
\usepackage{natbib}
\usepackage{hyperref}

\usepackage{orcidlink}
\DeclareUnicodeCharacter{FF0C}{,}

\newcommand\ep{\textit{EP}\xspace}

\newcommand\ixpe{\textit{IXPE}\xspace}

\newcommand\nicer{\textit{NICER}\xspace}

\newcommand\swift{\textit{Swift}\xspace}

\newcommand\targeta{1E~2259+586\xspace}
\newcommand\targetb{4U~0142+61\xspace}

\begin{document}

\title{A large radiatively loud anti-glitch in \targeta with a protracted recovery}

\author{
{Han-Long Peng}\inst{1}%
\and {Ming-Yu Ge}\inst{2, 3}\corrauth{gemy@ihep.ac.cn}%
\and {Shan-Shan Weng}\inst{1}\corrauth{wengss@njnu.edu.cn}%
\and {Roberto Taverna}\inst{4}%
\and {Shuang-Nan Zhang}\inst{2, 3}%
\and {Hua Feng}\inst{2, 3}%
\and {Yu-Jia Zheng}\inst{1}%
\and {Kai-Ge Chang}\inst{5}%
\and {Shi-Jie Zheng}\inst{2, 3}%
}

\institute{%
        School of Physics and Technology, Nanjing Normal University, Nanjing, 210023, Jiangsu, People's Republic of China
   \and Key Laboratory of Particle Astrophysics, Institute of High Energy Physics, Chinese Academy of Sciences, Beijing 100049, People's Republic of China
   \and University of Chinese Academy of Sciences, Chinese Academy of Sciences, Beijing 100049, People's Republic of China
   \and Department of Physics and Astronomy, University of Padova, Via Marzolo 8, Padova, I-35131, Italy
   \and Guangxi Key Laboratory for Relativistic Astrophysics, School of Physical Science and Technology, Guangxi University, Nanning 530004, People's Republic of China
}

   \date{}

\authorrunning{H. L. Peng et al.}

\abstract
{Anti-glitches are relatively rare and enigmatic timing anomalies. While recent high-cadence monitoring has revealed some radiatively quiet anti-glitches across various pulsar populations, the nature of radiatively loud anti-glitches---and their intricate connections to magnetospheric activities---remain poorly understood.}
{We aim to present a comprehensive observational characterization of a new, large anti-glitch in the magnetar \targeta, and to explore its phenomenological uniqueness relative to other magnetars.}
{We analyze the timing, flux, pulse profile evolution, and spectral properties of the source during this new event using multi-instrument X-ray observations from the \textit{EP}, \textit{NICER}, and \textit{IXPE}.}
{We report a new anti-glitch in \targeta occurring in 2025, which is the largest anti-glitch detected in this source, with $\Delta\nu \approx -1.174(7) \times 10^{-7}$\,Hz and $\Delta\nu/\nu \approx -8.19(4) \times 10^{-7}$. In stark contrast to the source's anti-glitch event in 2019, this anomaly was accompanied by extreme radiative activity. The persistent X-ray flux roughly doubled ($\sim 106\%$ increase), followed by a protracted recovery exceeding 320 days. Moreover, the pulse profile exhibited a modest but significant change: the depth of the valley between the main and secondary peaks decreased, and then recovered slowly along with the flux. In addition, the spectral properties changed across the event, and a short, soft 
$\gamma$-ray burst was detected by the \textit{Fermi}-GBM around the time of the event. These characteristics are similar to those observed during the radiatively loud anti-glitch in 2012.}
{\targeta is a unique source even among magnetars, exhibiting an anti-glitch rate comparable to that of normal spin-up glitches. Strikingly, its anti-glitch events can occur either accompanied by dramatic radiative changes or in relatively quiescent states. In addition, we find that the fractional amplitudes of its anti-glitches appear to increase monotonically; whether this trend persists merits further verification. These unique observational characteristics indicate that continuous, regular monitoring of \targeta is essential for understanding the underlying mechanisms of anti-glitches.}


\maketitle

\nolinenumbers

\section{Introduction}

Magnetars are a rare and extreme class of young, isolated neutron stars. Observationally, they are characterized by long spin periods ($P \sim 1$--$12\,\mathrm{s}$) and large spin-down rates ($\dot{P} \sim 10^{-13}$--$10^{-10}\,\mathrm{s\,\mathrm{s}^{-1}}$). These rotational parameters imply ultra-strong surface dipole magnetic fields ($B \sim 10^{14}$--$10^{15}\,\mathrm{G}$) and characteristic ages on the order of a few thousand years \citep{Olausen_2014}. Historically, they first appeared under the names of soft gamma-ray repeaters (SGRs) and anomalous X-ray pulsars (AXPs) \citep{Mazets_a, Fahlman_1981}.  Unlike ordinary radio pulsars, whose emission is powered by the loss of rotational energy, the generally accepted magnetar model posits that both the persistent and transient emissions of magnetars are primarily driven by the decay and instability of their ultra-strong magnetic fields \citep{Thompson_1995, Thompson_1996}. To date, approximately 30 confirmed and candidate magnetars have been cataloged, with a comprehensive and regularly updated listing available in the McGill Magnetar Catalog \footnote{\url{http://www.physics.mcgill.ca/~pulsar/magnetar/main.html}} \citep{Olausen_2014}.

Similar to canonical rotation-powered pulsars (RPPs), magnetars also experience glitches---sudden spin-up of their spin frequency \citep{Dib_2014}. Uniquely, however, magnetars appear to be the favored hosts for anti-glitches (sudden spin-down events) \citep{Archibald_2013}. Although growing observational data have recently identified this behavior in an accreting ultra-luminous X-ray pulsar (NGC 300 ULX-1) and some RPPs \citep{Ray_2019, Zhou_2024, Tuo_2024, Wang_2025}, magnetars remain the most prolific sources of anti-glitches. Moreover, unlike the radiatively quiet behavior typically associated with RPPs, magnetar anti-glitches are generally accompanied by energetic activities such as bursts and profound radiative variations \citep{Archibald_2017, Ge_2024}. Furthermore, as the only confirmed origin of fast radio bursts (FRBs) to date \citep{Andersen_2020, Bochenek_2020}, magnetars exhibit anti-glitches that appear to be intrinsically linked to FRB emissions \citep{Younes_2023}. However, the absolute scarcity of recorded magnetar anti-glitch events severely restricts our understanding of the complex internal structures and radiation mechanisms of these extreme objects.

\targeta, the first AXP identified, exhibits a spin period of approximately 7$\text{\,s}$ and is spatially associated with the supernova remnant (SNR) CTB 109 \citep{Fahlman_1981}. Based on its spin-down rate, the inferred surface dipole magnetic field strength is estimated to be $5.9 \times 10^{13}\text{\,G}$. In 2013, \citet{Archibald_2013} reported an anti-glitch in this source, which was accompanied by short X-ray bursts, a significant flux enhancement, and variations in both the first frequency derivative ($\dot{\nu}$) and the pulse profile. Additionally, using archival data from the \textit{Rossi X-ray Timing Explorer} (\textit{RXTE}), \citet{Dib_2014} and \citet{Icdem_2012} reported a prominent timing discontinuity that occurred in 2009. If fitted with a standard glitch model, this event similarly yields a negative frequency jump ($\Delta \nu < 0$). Furthermore, the 2009 event exhibited significant changes in the pulse profile (detected in an observation), X-ray flux (in two observations), and the frequency derivative \citep{Dib_2014}. Strikingly, in 2020, \citet{Younes_2020} reported a radiatively quiet anti-glitch from this source, during which neither the pulse profile nor the pulsed flux exhibited any significant alterations.

In this work, we report a large new anti-glitch in \targeta with a significant flux increase, based on recent observations from the Einstein Probe (\ep) and the Neutron star Interior Composition Explorer (\nicer), along with archival data from the Imaging X-ray Polarimetry Explorer (\ixpe). Notably, this is the largest anti-glitch detected in this source to date, featuring a prolonged relaxation phase exceeding 320 days. The structure of this paper is as follows: Section~\ref{sec:obs} describes the observations and the data reduction procedures. Section~\ref{sec:ana} presents the timing and flux analysis of \targeta. Finally, section \ref{sec:dis} provides the discussion and conclusions.

\section{Observation and data reduction \label{sec:obs}}

\subsection{EP/FXT}

The \ep, launched on 2024 January 9, is a space mission led by the Chinese Academy of Sciences \citep{Yuan_2022, Yuan_2025}. The observatory is equipped with two scientific instruments: the Wide-field X-ray Telescope (WXT) and the Follow-up X-ray Telescope (FXT). The EP/FXT \footnote{\url{https://epfxt.ihep.ac.cn/about}} is a Wolter-I focusing telescope operating in the 0.3--10 keV energy range, comprising two coaligned, identical units (FXT-A and FXT-B) \citep{Chen_2020}.

In this work, we employ 11 observations from the EP/FXT (see Table \ref{table:obs-used} for the detailed observation log). 
We process the EP-FXT observations with the \texttt{fxtchain} utility from the FXTDAS data analysis software package (version 1.30) \footnote{\url{http://epfxt.ihep.ac.cn/analysis}}. Running the \texttt{fxtchain} pipeline requires explicit assignment of input and output directory paths, alongside setting the \texttt{exitstage} keyword to 1. We apply the good time interval filtering criterion \(\mathrm{expr} = \mathrm{ELV}>5 \&\& \mathrm{COR}>6 \&\& \mathrm{SAA}==0 \&\& \mathrm{DYE\_ELV}>30\) to produce calibrated event files restricted to valid observation intervals. We visualize the preprocessed event files with \texttt{ds9}. A circular source region with a radius of \(30^{\prime\prime}\) is defined for the target source, while the background is extracted from an annular region with an inner radius of $200^{\prime\prime}$ and an outer radius of $400^{\prime\prime}$.
Source event files and spectra are extracted from cleaned event files via \texttt{Xselect}. We produce redistribution matrix files (RMFs) with \texttt{fxtrmfgen}, and enable PSF correction by assigning \texttt{psfcor=1} in the \texttt{fxtarfgen} task when building ancillary response files (ARFs). Barycentric corrections for the event files are performed via \texttt{fxtbary}.

\subsection{NICER} \label{sec:nicer data}

\nicer is an International Space Station payload designed for soft X-ray ($0.2\text{--}12\text{~keV}$) timing investigations of neutron stars \citep{Gendreau_2016}. Since its launch in 2017, \nicer has maintained regular, high-cadence monitoring of \targeta. In this work, we utilize all available observational data since 2023 June 12. Unfortunately, \nicer science observations have been suspended since June 17, 2025 \footnote{\url{https://heasarc.gsfc.nasa.gov/docs/nicer/index.html}}. As a result, we acquired only two observations following the anti-glitch event. 

All observational data are processed utilizing \texttt{HEASOFT} v6.36. We first use the default standard procedure of \texttt{nicerl2} pipeline (part of \texttt{NICERDAS} v15) to generate cleaned and calibrated event files. To eliminate contamination from particle-background flares, we then generate $12\text{--}15\text{\,keV}$ light curves with \texttt{nicerl3-lc} and apply the \texttt{maketime} tool to discard time bins where the count rate exceeds $0.2\text{ counts s}^{-1}$. The resulting GTI files are then incorporated by re-running \texttt{nicerl2}, yielding the final event files. Barycentric corrections are subsequently applied using the \texttt{barycorr} tool. Finally, we use \texttt{nicerl3-spect}, selecting the background model 3C50, to generate energy spectra and background products. For a more detailed description of the \nicer data processing pipeline, please refer to the webpage.\footnote{\url{https://heasarc.gsfc.nasa.gov/docs/nicer/nicer_analysis.html}}

\subsection{IXPE}

To investigate whether the X-ray flux has returned to its quiescent level, we also include two \ixpe observations of \targeta (see Table \ref{table:obs-used}). The analysis of the first \ixpe polarization results is presented in \citet{Heyl_2024} and \citet{Peng_2024}, while the polarization analysis of the new ToO observation and its comparison with the first observation will be reported in detail by the \ixpe collaboration.

We run the software \texttt{filter\_background.py} \footnote{\url{https://github.com/aledimarco/IXPE-background}} to filter clean particle events from level-2 event files \citep{Di_Marco_2023}.
Afterward, the \texttt{xpselect} tool of \texttt{ixpeobssim} (32.0.1) was adopted to define extraction regions \citep{Baldini_2022}: A circular region (radius of $60^{\prime\prime}$) and an annular region (inner and outer radii of $150^{\prime\prime}$ and $270^{\prime\prime}$) are selected for the extraction of source and background events, respectively.
Data from the three independent detector units (DUs) are combined to improve data quality for the timing analysis, and the barycentric corrections are also applied using the \texttt{barycorr} tool. For a more detailed data processing regarding \ixpe, readers can refer to these articles \citep{Baldini_2022, Zane_2023, Di_Marco_2023} and this webpage.\footnote{\url{https://github.com/lucabaldini/ixpeobssim}}

\begin{table*}
\caption{The observations of \targeta used in this work.}
\label{table:obs-used}
\centering
\begin{tabular}{ccccc}
\hline
Telescope & Date (UTC)        & ObsID           & MJD      & Exposure \\
          & YYYY-MM-DD              &                 &          & s       \\
\hline
          & 2024-07-25         & 11908462848     & 60516.7         & 3158.8 \\
          & 2024-08-21         & 11908522753     & 60543.6         & 1418.9 \\
          & 2024-09-17         & 11908647686     & 60570.6         & 1127.3 \\
          & 2024-10-15         & 11908722953     & 60598.5         & 1133.9 \\
          & 2024-11-12         & 11908828930     & 60626.9         & 1305.0 \\
EP-FXT (11)    & 2025-08-17         & 11900351744     & 60904.7         & 1259.9 \\
          & 2025-10-13         & 11900434176     & 60961.6         & 1198.9 \\
          & 2025-11-03         & 11900460672     & 60982.9         & 1280.9 \\
          & 2025-11-22         & 11900486016     & 61001.8         & 1258.9 \\
          & 2025-12-11         & 11900509952     & 61020.3         & 1431.9\\
          & 2025-12-29         & 11900532608     & 61038.0         & 1450.6 \\
\hline
          & 2023-06-12         & 6533040701      & 60107.5         &  \\
NICER (24)     & to             & ...             &  to             & 18549.0  \\
          & 2025-06-13         & 8598040801      & 60839.0         &  \\
\hline
IXPE (2)      & 2023-06-02         & 02007899     & 60097.9         & 328530.1 \\
             & 2026-04-24         & 05250401     & 61154.3         & 395952.1 \\
\hline
\end{tabular}
\end{table*}


\section{Results} \label{sec:ana}

\subsection{Timing \label{sec:ana-1}}

\begin{figure*}
\begin{center}
\includegraphics[width=0.7\textwidth,clip]{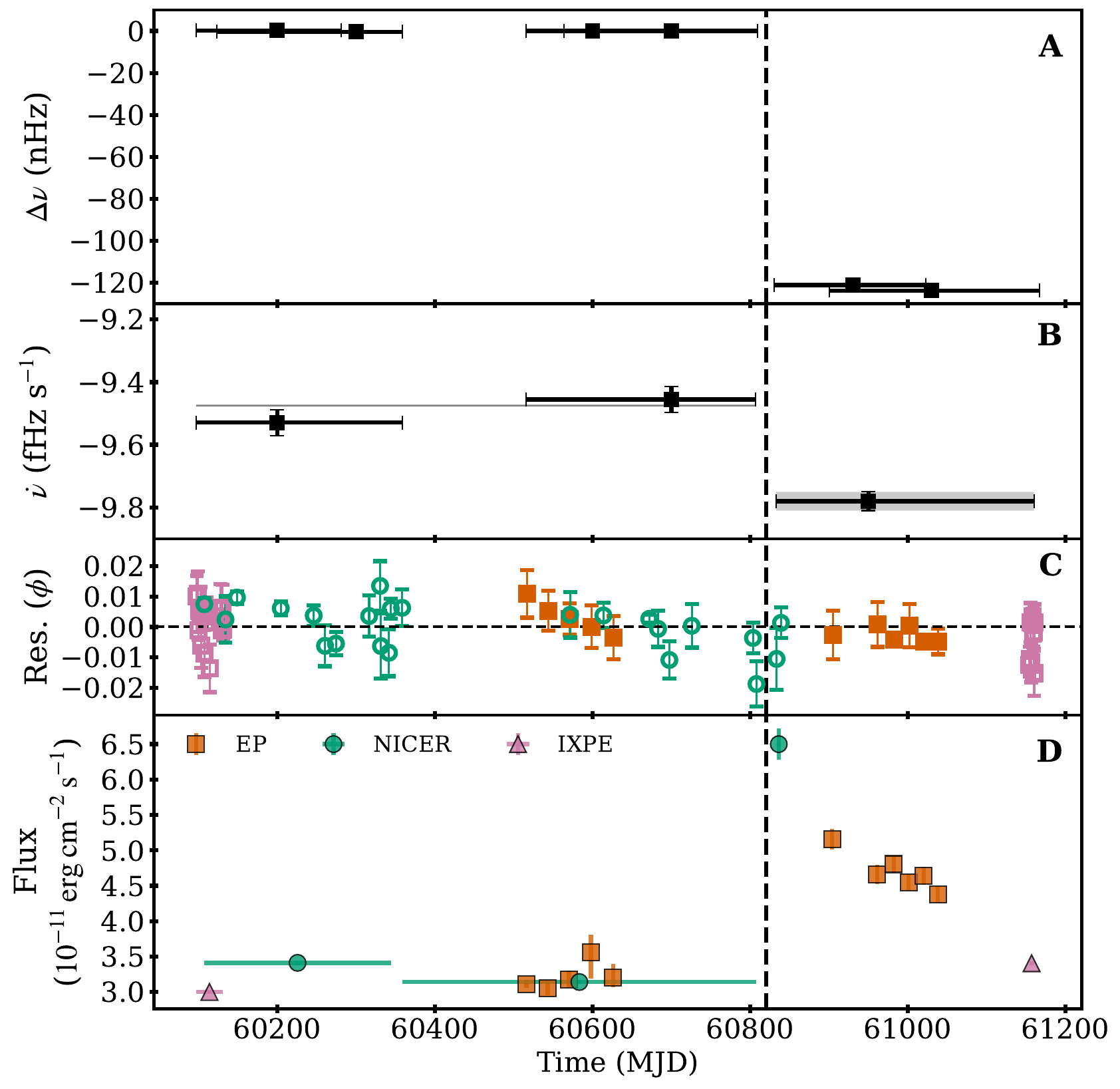}
\caption{Spin and flux evolution of \targeta. The vertical dashed line marks the epoch of the anti-glitch. 
(A) Frequency evolution as a function of time, with a long-term linear trend subtracted.
(B) Frequency derivative ($\dot{\nu}$) as a function of time. The gray shaded regions indicate the $\dot{\nu}$ values and their associated uncertainties obtained from fits to all ToAs within each of the two ephemeris segments (S1 and S2 in Table~\ref{table:timing-1}).
(C) Timing residuals for all ToAs. (D) Flux evolution as a function of time. Data points from \textit{EP-FXT}, \textit{IXPE}, and \textit{NICER} are shown in orange-red, reddish-purple, and bluish-green, respectively.} 
\label{timing-flux}
\end{center}
\end{figure*}

We first adopt the ephemerides provided by \citet{Peng_2026}, which describe the spin evolution of \targeta from 2017 to 2024. The times of arrival (ToAs) and their uncertainties are calculated by cross-correlating each pulse profile with a cumulative high signal-to-noise profile template \citep{Taylor_1992, Tuo_2024}. In the absence of new timing anomalies, this ephemeris adequately models the derived ToAs, which manifests as featureless residuals in \texttt{TEMPO2} \citep{Hobbs_2006}. Conversely, the onset of an anomaly introduces significant systematic trends into the residuals. We then re-perform the timing analysis to extract the post-anomaly ToAs and construct a new ephemeris with \texttt{TEMPO2} \citep{Ge_2012, Dib_2014}. To trace the spin evolution of the target, we employ a partially phase-coherent timing (PPCT) analysis \citep{Ferdman_2015, Livingstone_2005}, dividing the ToAs into segments and fitting them separately using \texttt{TEMPO2}. The resulting spin frequency and its derivative are shown in Figure~\ref{timing-flux}.

As shown in the top panel of Figure~\ref{timing-flux}, a prominent change in the spin frequency occurs around MJD 60820. Although the change in frequency derivative is not as prominent as that in frequency, it still shows a small increase of approximately $3 \times 10^{-16}$\,Hz $s^{-1}$ compared to the pre-antiglitch level (see the second panel of Figure~\ref{timing-flux}). Following the initial PPCT results, we apply fully phase-coherent timing (FPCT) analysis---globally fitting all the ToAs---to derive more accurate and continuous timing solutions \citep{Livingstone_2005, Peng_2026}. The values for each parameter are listed in Table~\ref{table:timing-1} and the corresponding timing residuals are displayed in the third panel of Figure~\ref{timing-flux}. Compared to previously reported events \citep{Icdem_2012, Dib_2014, Archibald_2013, Younes_2020}, the magnitude of the frequency change ($\Delta\nu = -1.174(7) \times 10^{-7}\,\mathrm{Hz}$) and the fractional frequency change ($\Delta\nu/\nu = -8.19(4) \times 10^{-7}$) represent the largest anti-glitch ever detected in \targeta, with an rms residual of 21.4\,ms. Note that throughout this text, all spin-parameter solutions obtained with \texttt{TEMPO2} and fitting results from \texttt{XSPEC} are reported at the 1\,$\sigma$ confidence level.

\begin{table*}
\caption{Spin parameters of \targeta.}
\label{table:timing-1}
\centering
\begin{tabular}{lccc}
\hline
Parameter                                & S1               & S2                 & G                  \\
\hline
MJD range                                & 60000--60810     & 60830--61170       & 60000--61170       \\
Epoch (MJD)                              & 60700            & 60950              & 60700              \\
$\nu$ ($\mathrm{Hz}$)                    & 0.1432807289(1)  & 0.1432804034(3)    & 0.1432807289(1)    \\
$\dot\nu$ ($\mathrm{fHz\ s^{-1}}$)       & $-9.475(6)$      & $-9.78(3)$         & $-9.475(6)$        \\
\hline
$t_{g}$ (MJD)                            & -          & -            & 60820              \\
$\triangle\phi_{g}$                      & -          & -            & $-0.013(6)$        \\
$\triangle\nu_{g}$ ($10^{-7}\ \mathrm{Hz}$) & -       & -            & $-1.174(7)$        \\
$\triangle\dot\nu_{g}$ ($10^{-16}\ \mathrm{Hz\ s^{-1}}$) & - & -    & $-3.0(3)$          \\
$\triangle\nu_{g}/\nu_{g}$ ($10^{-7}$)   & -          & -            & $-8.19(4)$         \\  
$\triangle\dot\nu_{g}/\dot\nu_{g}$       & -          & -            & 0.032(3)           \\
\hline
Residuals (ms)                           & 21.943           & 20.204             & 21.408             \\
\hline
\end{tabular}
\tablefoot{S1 and S2 refer to the pre- and post-event timing models, respectively, while G denotes the global ephemeris encompassing the anti-glitch parameters.}
\end{table*}

\begin{figure*}
\begin{center}
\includegraphics[width=0.7\textwidth,clip]{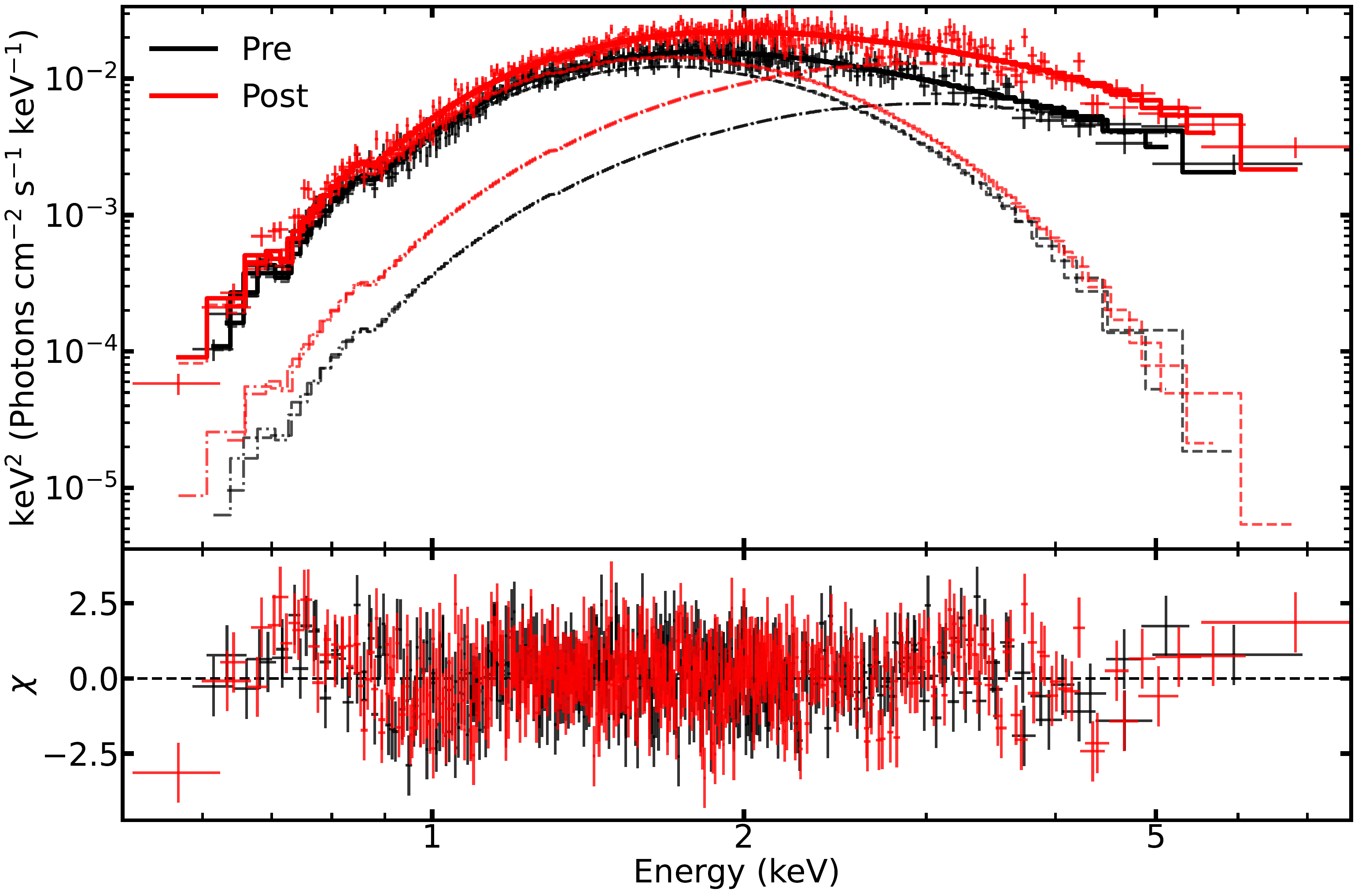}
\caption{Phase-averaged spectrum of \targeta before and after the anti-glitch (top panel) together with the corresponding fit residuals (bottom panel). The pre- (black) and post-antiglitch (red) spectra are shown by points with error bars, while their corresponding best-fit total models (solid lines) and individual components (dashed lines) are also overplotted.} 
\label{phase_averaged}
\end{center}
\end{figure*}

\begin{figure*}
\begin{center}
\includegraphics[width=0.95\textwidth,clip]{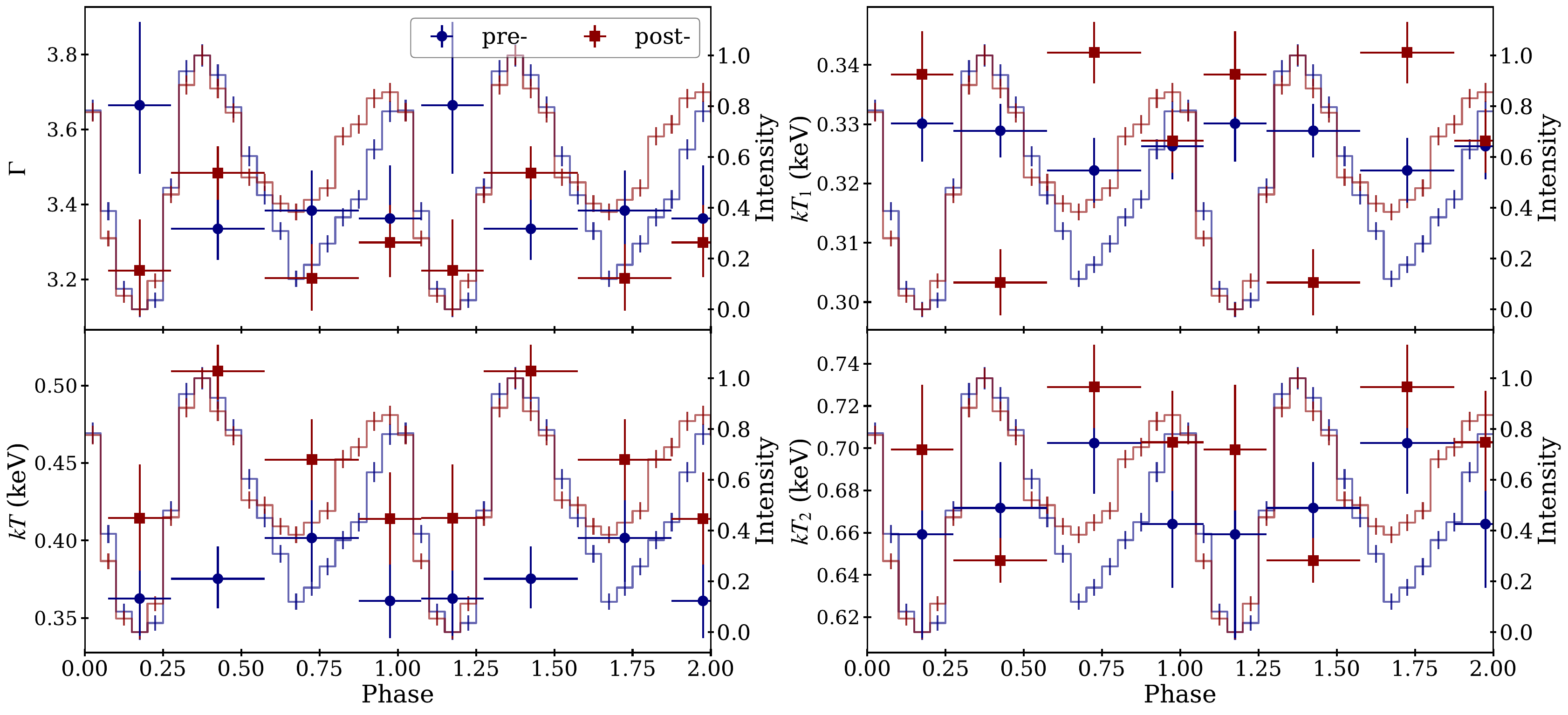}
\caption{Phase-dependent evolution of the spectral parameters for \targeta. The points with error bars show the phase-resolved spectral parameter values, where blue and red correspond to the pre- and post-anti-glitch results, respectively. Horizontal error bars indicate the phase-width interval adopted in the phase-resolved analysis; the parameter values and uncertainties are derived from XSPEC fits at the $1\sigma$ confidence level. The left column displays the results derived from the blackbody plus power-law model, while the right column shows those from the double blackbody model. The average pulse profiles before (blue) and after (red) the anti-glitch are also shown in the four panels.} 
\label{phase_resolve}
\end{center}
\end{figure*}

\begin{figure}
\begin{center}
\includegraphics[width=0.48\textwidth,clip]{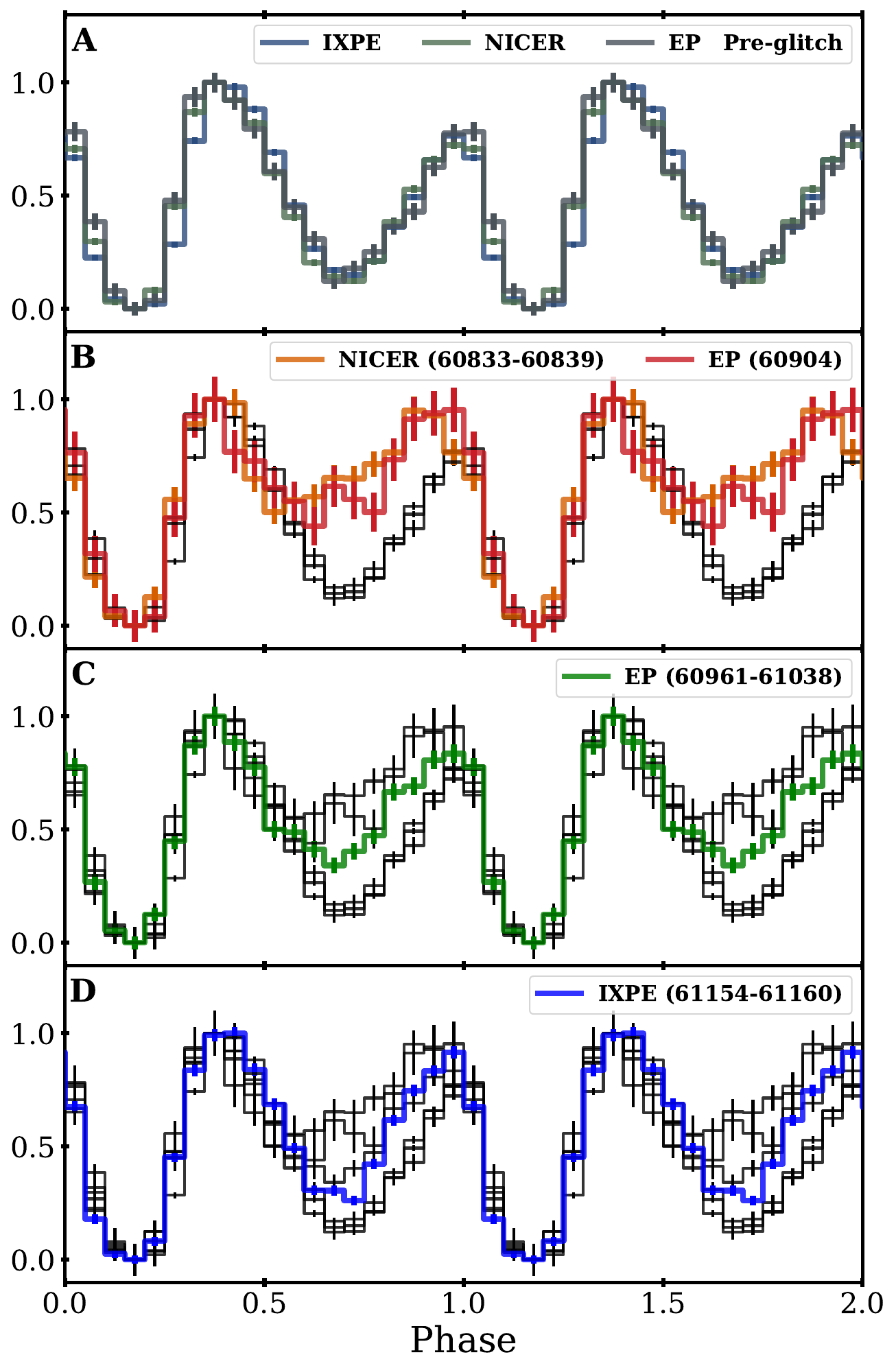}
\caption{Pulse profile evolution of \targeta. (A): Consistent pre-antiglitch pulse profiles obtained by three different detectors. (B): The earliest post-antiglitch pulse profiles from \nicer (orange) and \ep (red). The \nicer data comprise two observations spanning MJD 60833--60839, while the \ep profile is derived from a single observation on MJD 60904. (C): Same as Panel (B), with the addition of pulse profiles from five subsequent \ep observations covering MJD 60961--61038. (D): Further incorporating the \ixpe observation spanning MJD 61154--61160. In each panel, the newly added profiles are highlighted in color, while previously shown profiles are rendered in black.} 
\label{pro}
\end{center}
\end{figure}

\subsection{Flux evolution} \label{sec:ana-2}

For the spectral analysis, we utilize \texttt{XSPEC} to fit all individual \ep and \ixpe observations separately \citep{Arnaud_1996}. For the \nicer data, we employ the \texttt{addspec} tool to combine the pre-antiglitch observations into two high-quality spectra. The corresponding response files are also merged using \texttt{addrmf} and \texttt{addarf} accordingly. We fit all spectra using two models commonly applied to magnetars: a double blackbody or a blackbody plus a power-law \citep{Mereghetti_2008, Turolla_2015}. Notably, after applying our filtering criteria (see Section~\ref{sec:nicer data}), the merged post-anti-glitch \nicer spectrum has a total exposure time of only $73\text{\,s}$. Given the low exposure, we find that the best-fitting model is a single blackbody one, while a multi-component fit would leave the additional component unconstrained. Moreover, the flux from the single blackbody model is approximately twice the pre-antiglitch level, consistent with the observed count-rate doubling. Panel D of Figure~\ref{timing-flux} illustrates the evolution of the absorbed flux, derived using the double blackbody model. Similarly, adopting a blackbody plus power-law model yields a consistent absorbed flux evolution.

As shown in panel (D) of Figure~\ref{timing-flux}, the X-ray flux of \targeta exhibits a prominent increase around MJD 60820, subsequently entering a long-term decay phase. Specifically, the flux increases by approximately $106\%$ for \nicer and $\sim60\%$ for \ep. After nearly a year, the \ixpe data indicate that a residual flux enhancement of $\sim13\%$ still persists. Based on the time span between the first post-antiglitch \nicer observation and the latest \ixpe epoch, we deduce that after approximately 320 days of decay, the flux has not yet returned to its quiescent level. It is important to note that we perform joint spectral fits of the \nicer data with those from \ep and \ixpe separately. By dividing the \ep and \ixpe fluxes by their respective cross-calibration constants, we construct a visually continuous flux evolution picture. However, owing to the inherent differences in instrument responses, energy bands, and effective areas among the three telescopes, we refrain from combining their flux measurements to perform a global exponential decay fit (or any other empirical modeling). Instead, we independently calculate the fractional flux increase for each instrument and utilize the respective observation epochs to estimate the decay timescale.

For the \ep observations, we merge the pre- and post-antiglitch data following the same procedure applied to the \nicer spectra. Subsequently, we perform both phase-averaged and phase-resolved spectral analyses on the merged datasets. For the phase-averaged spectra, fitting with an absorbed power-law plus blackbody model (\texttt{constant*tbabs*(powerlaw + bbodyrad)}) reveals that following the anti-glitch, the photon index increases and the blackbody temperature rises, indicating a softer non-thermal component and hotter thermal emission. Alternatively, fitting with a double blackbody model (\texttt{constant*tbabs*(bbodyrad + bbodyrad)}) shows that while the temperatures of the two thermal components do not exhibit a significant increase, the emitting radius of the hotter blackbody expands. Regardless of the applied model, a direct comparison between the pre- and post-antiglitch spectra reveals that the flux enhancement primarily occurs above 2\,keV, as illustrated in Figure \ref{phase_averaged} (using the double blackbody model as an example). Notably, given the time gap of over six months between the merged pre- and post-glitch observations, we leave the hydrogen column density ($N_{\rm H}$) as a free parameter rather than tying it across the two epochs. 

For the phase-resolved spectra, fitting with the PL + BB model reveals that the spectrum in all phase intervals, aside from the main peak, becomes harder, accompanied by a significant global increase in temperature. Furthermore, the maximum values of both the photon index and the blackbody temperature shift to the phase of the main peak. Alternatively, fits with the double BB model reveal a clear shift in the temperature distributions of both blackbody components: they are uniform before the anti-glitch, but become anti-correlated with the pulse profile afterward. The phase-dependent evolution of the spectral fit parameters is shown in Figure \ref{phase_resolve}.

Finally, we search for short bursts in the \ep and \nicer observations but detect no significant events across various time bins. Furthermore, \citet{Chu_2026} systematically searches the \nicer data for magnetar bursts using the Bayesian block algorithm. They similarly report an absence of short bursts around the epoch of this anti-glitch. However, a recent systematic search for magnetar short bursts in \textit{Fermi}-GBM data by \citet{Godwin_2026} reveals a 31~ms burst around MJD 60817. With a flux of approximately $3 \times 10^{-6} \mathrm{erg\,cm^{-2}\,s^{-1}}$, this event is only half as bright as the burst recorded in 2012 \citep{Archibald_2013}. This relative faintness may explain why it fails to trigger those detectors.

\subsection{Pulse profile evolution} \label{sec:ana-3}

We also analyze the pulse profile evolution of \targeta. We divide the data into seven distinct epochs, corresponding to the pre- and post-antiglitch cumulative profiles from \ixpe, \nicer, and \ep. Notably, because the first post-antiglitch \ep observation occurs between the two \nicer epochs and the remaining five \ep epochs, we fold its pulse profile independently to better trace the morphological evolution. 

We observe an overall increase in the profile count rate, which is consistent with the observed global flux enhancement. Furthermore, to compare the profile shapes directly, we normalize all pulse profiles to a range of 0 to 1 using $C_{\text{norm}} = \frac{C - C_{\min}}{C_{\max} - C_{\min}}$. As shown in Figure~\ref{pro}, the depth of the dip between the main and secondary peaks decreases significantly following the anti-glitch, before gradually recovering over time. As of the latest \ixpe observation, the pulse profile has almost recovered but a slight increase remains compared to the pre-antiglitch baseline. This morphological state is also in agreement with the 13\% flux increase still present in the \ixpe flux (see Figure \ref{timing-flux}).

\section{Discussion and summary} \label{sec:dis}

In this work, we comprehensively analyze the \ep observations in conjunction with archival \nicer and \ixpe data. We find that \targeta underwent a large anti-glitch around MJD 60820, with $\Delta\nu = -1.174\times10^{-7}$ Hz ($\Delta\nu/\nu = -8.19\times10^{-7}$). This event was accompanied by a significant X-ray flux enhancement, variations in the pulse profile, and a protracted decay timescale exceeding 320 days. Furthermore, the properties of both the phase-averaged and phase-resolved spectra underwent changes. 
Next, we compare these findings with the historical behavior of \targeta, and subsequently compare this source with other objects that have experienced anti-glitches.

\subsection{Comparison with historical anti-glitches in \targeta}
\begin{figure}
\begin{center}
\includegraphics[width=0.5\textwidth,clip]{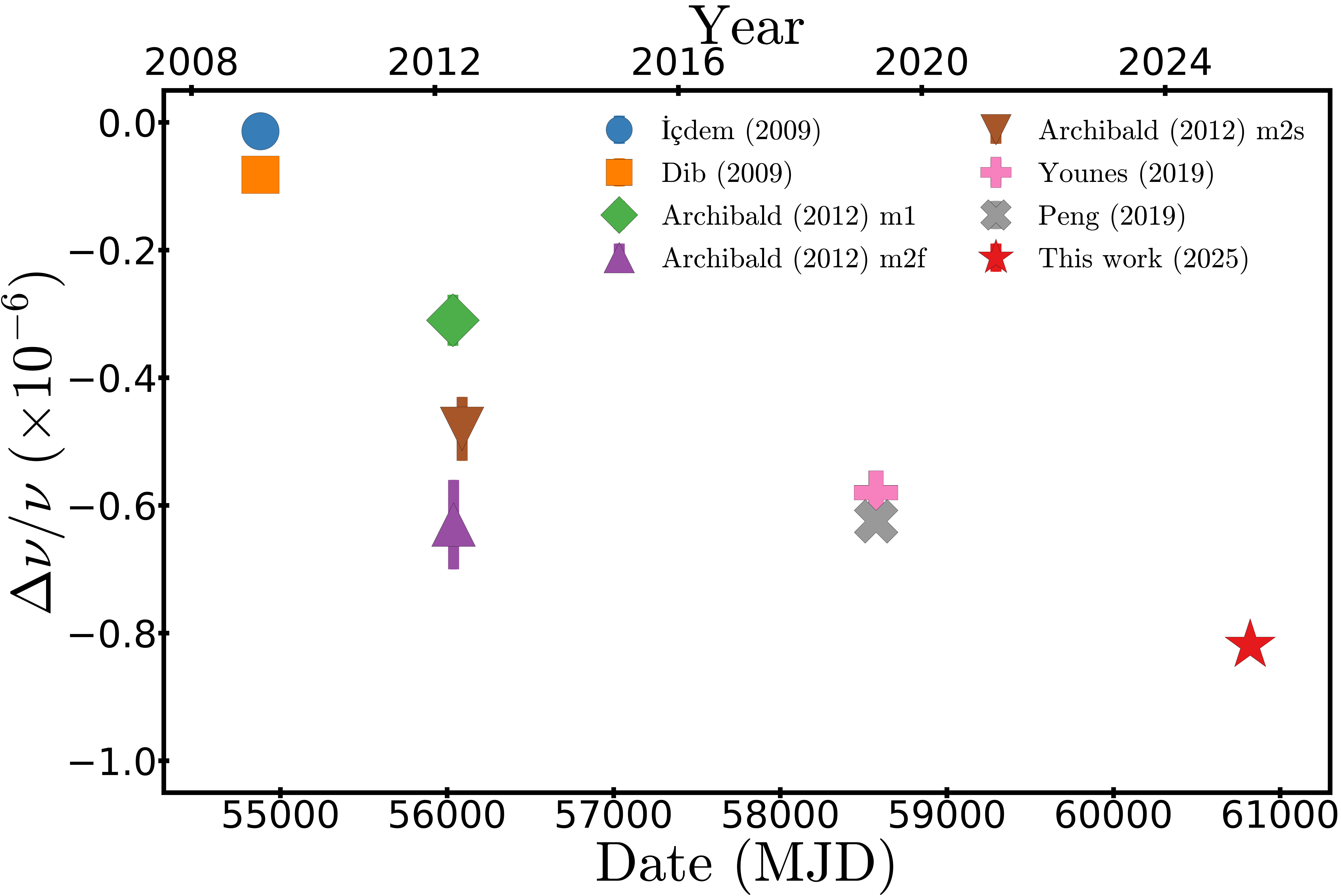}
\caption{Fractional amplitudes of the historical anti-glitches in \targeta as a function of time. The top x-axis indicates the corresponding observation years. Different symbols and colors represent various timing solutions: blue circles and orange squares denote the solutions for the 2009 anti-glitch by \cite{Icdem_2012} and \cite{Dib_2014}, respectively. The green diamond represents the first model of the 2012 event by \cite{Archibald_2013}, while the purple upward triangles and brown downward triangles indicate the first and second events of their second model, respectively. The pink filled-plus and grey crosses represent the timing solutions for the 2019 event by \cite{Younes_2020} and \cite{Peng_2026}. The red star indicates the result from this work. Error bars are shown for all data points.}
\label{2259-his}
\end{center}
\end{figure}

Prior to this event, \targeta had experienced an anti-glitch in 2012 ($\Delta\nu = -4.5\times10^{-8}$ Hz, $\Delta\nu/\nu = -3.1\times10^{-7}$), another in 2019 ($\Delta\nu = -8.3\times10^{-8}$ Hz, $\Delta\nu/\nu = -5.8\times10^{-7}$), and a candidate event in 2009 ($\Delta\nu = -1.2 \times 10^{-8}$\,Hz, $\Delta\nu/\nu = -8.2 \times 10^{-8}$) \citep{Icdem_2012, Archibald_2013, Dib_2014, Younes_2020, Peng_2026}. Compared with these historical records, the newly detected 2025 anomaly represents the largest anti-glitch detected in this source to date (also see Figure \ref{2259-his}).

Interestingly, plotting these anti-glitch events reveals a monotonic trend: the fractional amplitudes ($\Delta\nu/\nu$) of \targeta appear to increase with time (as shown in Figure \ref{2259-his}). However, owing to the limited sample, whether this trend persists remains to be verified with future observations. We also analyzed the $\triangle\dot\nu/\dot\nu$ values and found no correlation. Finally, no correlation was found when glitches and anti-glitches were analyzed together, which may suggest a difference between the two classes.

Phenomenologically, the 2025 anti-glitch exhibits striking similarities to the 2012 radiatively loud anti-glitch \citep{Archibald_2013}. Both events were accompanied by soft $\gamma$-ray bursts detected around the timing epoch, a persistent X-ray flux doubling, and a remarkable morphological change in the pulse profile. Moreover, the enhanced flux and the altered pulse profile in both cases recovered slowly and synchronously over a protracted timescale. This remarkable recurrence suggests that radiatively loud anti-glitches in \targeta may originate from a repeatable physical process.

Conversely, the 2025 event stands in sharp contrast to the 2019 anti-glitch reported by \cite{Younes_2020}. The 2019 event was a relatively radiation-quiet one, showing no detectable changes in pulsed flux or pulse profile, and no associated short bursts \citep{Younes_2020, Peng_2026}. This remarkable coexistence of diverse behaviors within a single source poses a significant challenge to unified theoretical models, implying that the physical triggers of magnetar spin-down glitches may be variable. While radiatively quiet events might be primarily driven by internal mechanisms---such as angular momentum exchange via vortex inward drift or superfluid partial conversion \citep{Zhou_2024, Tuo_2024}---the radiatively loud events strongly indicate the active participation of external mechanisms, such as magnetic field line twisting and particle wind braking torques \citep{Tong_2014, Younes_2023}. In particular, the significant pulse profile reshaping, the phase-dependent evolution of spectral parameters, and the exceptionally long decay timescale reported here point toward a profound and persistent reconfiguration of the surface hot spots and the external magnetic field topology.

\subsection{Comparison with other magnetars and non-magnetar sources}
In recent years, high-cadence timing programs have unveiled anti-glitch behaviors in non-magnetar populations, including canonical RPPs such as PSR J1522$-$5735 \citep{Zhou_2024}, PSR B0540$-$69 \citep{Tuo_2024}, and PSR J1835$-$1106 \citep{Wang_2025}, as well as in the accreting ultraluminous X-ray pulsar NGC 300 ULX-1 \citep{Ray_2019}. However, profound phenomenological gaps remain.

The primary distinction lies in the radiation response. All anti-glitches identified in RPPs to date are strictly radiatively quiet; high-sensitivity multi-wavelength monitoring has shown no detectable changes in flux, pulse profile, or spectral shape across those events \citep{Tuo_2024, Zhou_2024, Wang_2025}. In contrast, magnetar anti-glitches almost always exhibit dramatic radiative variability. Additionally, anti-glitches in RPPs are generally small in magnitude ($\vert\Delta\nu/\nu\vert \sim 10^{-9}$ to $10^{-8}$), whereas radiatively loud magnetar anti-glitches can reach much larger values, even up to $10^{-6}$ or more \citep{Ge_2024, Younes_2023, Archibald_2017, Woods_1999}. It remains an open question whether the ultra-strong magnetic field itself, distinct internal core-crust dynamics, or some other factors dictate this stark contrast.

A secondary aspect is the occurrence rate. Magnetars account for the vast majority of the current anti-glitch sample. Yet, even within the magnetar population, the distribution is highly uneven. While most magnetars have exhibited only one or two such anomalies, \targeta and \targetb show a remarkably high anti-glitch occurrence rate \citep{Younes_2020, Peng_2026}, which is comparable to their normal spin-up glitch rates. What causes the occurrence rates of the two to be so different from those of other magnetars remains a mystery.

The exceptionally high anti-glitch rate of \targeta, coupled with its unique tendency to exhibit anti-glitches in both dramatic bursting states and relatively burstless states, firmly establishes it as an unparalleled astrophysical laboratory. Regular and intensive monitoring of this specific source is crucial to better understand the causes of these distinctive events. Furthermore, continued systematic monitoring of the broader magnetar population is also essential. Unfortunately, no further observations from \nicer or \swift will be available in the future. However, our results have shown that \ep can take over the task of magnetar timing observations. 
Moreover, China plans to launch the enhanced X-ray Timing and Polarimetry mission (eXTP) in the coming years \citep{Zhang_2025}, which will be capable of performing simultaneous imaging, timing, spectral, and polarimetric analyses of sources \citep{Ge_2025}. Its unique and comprehensive observational capabilities are poised to make profound contributions to our understanding of magnetar physics.

\begin{acknowledgements}
This work is based on observations made with the \ep mission and software from the Einstein Probe Science Center (EPSC), as well as data obtained by the \nicer and \ixpe missions. The authors acknowledge support from the National Natural Science Foundation of China (Grants 12473041, 12373051 and 12673057) and China's Space Origins Exploration Program. H.-L.P. acknowledges financial support from the Postgraduate Research \& Practice Innovation Program of Jiangsu Province (181200003026176). This work is also supported by the National Key R\&D Program of China (2021YFA0718500) from the Minister of Science and Technology of China (MOST).
\end{acknowledgements}

\bibliographystyle{aa} 
\bibliography{main}{} 

\end{document}